\documentclass[onecolumn,,preprint]{revtex4-1}
\usepackage{amssymb}
\usepackage{amsfonts}
\usepackage{amsmath}
\usepackage{graphicx}
\usepackage{epstopdf}
\usepackage{float}

\begin{document}

\title{Relativistic quantum motion of spin-0 particles under the influence
of non-inertial effects in the cosmic string space-time}
\author{L. C. N. Santos}
\author{C. C. Barros Jr.}

\begin{abstract}
We study solutions for the Klein--Gordon equation with vector and scalar
potentials of the Coulomb types under the influence of non-inertial effects 
in the space-time of topological defects. We also investigate a quantum
particle described by the Klein--Gordon oscillator in the background
space-time generated by a string. An important result obtained is that the
non-inertial effects restrict the physical region of the space-time where the
particle can be placed. In addition, we show that these potentials can form
bound states for the relativistic wave equation equation in this kind of
background.
\end{abstract}

\maketitle

\preprint{HEP/123-qed}

\affiliation{Depto de F\'{\i}sica - CFM - Universidade Federal de Santa Catarina, CP. 476
- CEP 88.040 - 900, Florian\'{o}polis - SC - Brazil}

\volumeyear{} \volumenumber{} \issuenumber{} \eid{identifier} \startpage{1} %
\endpage{10}

\section{\protect\bigskip introduction}

The quantum field theory (QFT) in curved space-time can be considered as a
first approximation to quantum gravity. Moreover, to make a consistent
quantum field theory in a gravitational background, it is necessary to
analyze the single particle states, in this way, efforts have been applied
in order to find an adequate formulation of the relativistic equation of
motion for particles in a curved space-time. In recent years, there has been
a significant increase of interest in the study of gravitational effects on
quantum mechanical systems (single particle states) \cite%
{Castro1,Castro2,incert1,parker1,hcurvo1,Barros1,barros3,string8}. In addition, the
physics of a neutrino in a curved metric is considered in \cite{neutrino1}
by Wheeler and Brill who presented a detailed analysis of the interaction of
neutrinos and gravitational fields.

Topological defects are other kind of system that may be studied with
this purpose. These defects are intriguing systems, that are supposed to be
created during a symmetry breaking phase transition in the early universe 
\cite{string1,string2,string7} and may be
considered as topological defects in the space-time structure predicted by a
large class of theories. These structures are candidates for the generation
of observable astrophysical phenomena such as high energy cosmic rays, gamma
ray burst and gravitational waves \cite{string11}. The recent discovery of
gravitational waves by the LIGO collaboration \cite{waves1} suggests that a
promising way to detect cosmic strings is to search for the
gravitational-wave radiation they would produce.

A key feature is that the geometry around them is locally flat, but this is
not a global feature. In \cite{santos1}, the effects of magnetic
fields in the metric have been considered. A relativistic wave equation for
spin 1/2 particles in the Melvin space-time, a space-time where the metric
is determined by a magnetic field, has been obtained and the effects of very
intense magnetic fields in the energy levels, as intense as the ones
expected to be produced in ultra-relativistic heavy-ion collisions, has been
investigated.

Non-inertial effects on physical systems is another kind of aspect that have
been studied in many works in the literature \cite%
{santos3,inercial12,inercial13,inercial14,inercial15,inercial16,inercial17,inercial18,inercial19}%
. A special case of a non-inertial system is the rotating reference system.
In \cite{bakke2,bakke3}, a rotational reference system in the Minkowski space-time, is
investigate. Notably, in those papers it was shown that the geometry of the
space-time can play the role of a \ hard-wall potential. Another example of
a non-inertial system is the Mashhoon effect, that is the coupling of the
spin of the particles with the angular velocity of the rotating reference system and
it arises from the influence of these non-inertial frames when interference
effects are considered \cite{inercial4}.

In this contribution, we will study scalar bosons in a cosmic string
space-time by considering the relativistic wave equation with a vector
potential $v(r)=\kappa /r$ and a scalar potential $s\left( r\right) =\eta /r$%
, where $r$ is the radial coordinate with $\eta $ and $\kappa $ constants.
Afterwards, we will examine a similar problem, the Klein--Gordon oscillator
inside a topological defect space-time. Moreover, a rotational reference system in the
conical space-time will be considered in both cases, and we will show that
non-inertial effects reduce the physical region of the space-time where the
quantum particles can be placed, and furthermore the energy levels are shifted by the
non-inertial effects on the particle. This feature is an
indicator of a nontrivial phenomenon: the coupling between the angular
quantum number and the angular velocity of the rotating reference system. Afterwards,
we will show that these potentials can form bound states for the spin-0
equation in this space-time. This paper is part of a study where we are
interested in making a systematic exploration of the properties of quantum
systems inside spaces with different kinds of structures \cite%
{santos1,santos2}.

The paper is organized as follows: In Section II, we will describe the the
space-time of topological defect and the transformation from space-time
coordinates to rotating coordinates. In Section III, the relativistic wave
equation with vector and scalar potentials of the Coulomb types in the
space-time of topological defects will be determined and in Section IV the
Klein--Gordon oscillator will be solved. Finally, Section V presents our
conclusions. In this work, we use natural units where $c=G=\hbar =1.$

\section{The cosmic string and the non-inertial reference frame}

In this section we will describe the relationship between the metric of a
topological defect and the effects of the rotation of a reference frame. The
cosmic string space-time is a solution of Einstein's field equations and it
describes a space-time determined by an infinitely long straight string. The
string space-time is assumed to be static and cylindrically symmetric, and
then, the distance element re\-pre\-sen\-ting this system can be written in
the form \cite{string5,Castro2}%
\begin{equation}
ds^{2}=-dt^{\prime 2}+dr^{\prime 2}+\alpha ^{2}r^{\prime 2}d\phi ^{\prime
2}+dz^{\prime 2},  \label{eq1}
\end{equation}%
where $\alpha =1-4G\mu $ and $\mu $ is the mass density of the string. In
this space-time, the coordinates range is represented in the following way:
the azimuthal angle range is $\phi ^{\prime }\in \lbrack 0,2\pi )$ while $%
r^{\prime }$ and $z^{\prime }$ are $r^{\prime }$ $\in \lbrack 0,\infty )$
and $z^{\prime }\in (-\infty ,\infty )$ respectively. The parameter $\alpha $
is related to the curvature of space-time. It may assume values in which $%
\alpha \leq 1$ or $\alpha >1$, and in this case, it corresponds to a
space-time of topological defect with negative curvature. In this work, we
are interested in studying the case $0<\alpha <1$. The transformation of the
metric $\left( \ref{eq1}\right) $ for a rotational reference system may be made
by considering a coordinate transformation \cite{bakke4,bakke5}%
\begin{equation}
t^{\prime }=t,\text{ \ \ }r^{\prime }=r,\text{ \ \ \ }\phi ^{\prime }=\phi
+\omega t,\text{ \ \ }z^{\prime }=z,  \label{eq2}
\end{equation}%
where $\omega $ is angular velocity of the rotational reference system, which we assume
to be positive. In\-ser\-ting this transformation into eq. (\ref{eq1}) we
obtain the line element%
\begin{align}
ds^{2}& =-\left( 1-\alpha ^{2}r^{2}\omega ^{2}\right) dt^{2}+2\alpha
^{2}r^{2}\omega dtd\phi   \notag \\
& +dr^{2}+\alpha ^{2}r^{2}d\phi ^{2}+dz^{2},  \label{eq3}
\end{align}%
that may be associated with the covariant metric tensor 
\begin{equation}
g_{\mu \eta }=\left( 
\begin{array}{cccc}
-\left( 1-\alpha ^{2}r^{2}\omega ^{2}\right)  & 0 & 0 & \alpha
^{2}r^{2}\omega  \\ 
0 & 1 & 0 & 0 \\ 
0 & 0 & 1 & 0 \\ 
\alpha ^{2}r^{2}\omega  & 0 & 0 & \alpha ^{2}r^{2}%
\end{array}%
\right) .  \label{eq3b}
\end{equation}%
It is possible to see that  $g_{\mu \eta }$ is a non-diagonal metric tensor where
effects of the topology of the space and the rotation of the reference system are taken
into account. An interesting feature of the equation (\ref{eq3}) is the
condition 
\begin{equation}
0<r<1/\alpha \omega .  \label{eq4}
\end{equation}%
This is related to the result that for $r>$ $1/\alpha \omega $ the
velocity of  particles is greater than the velocity of the light, for this
reason, is convenient to restrict $r$ to the range (0,$1/\alpha \omega $%
). Hence, the radial wave function  must vanish at $r=1/\alpha
\omega $ and consequently the system presents two classes of
solutions that depend on the value of the product $\alpha \omega $. Thus the
first case is obtained by adopting the limit $\alpha \omega \ll 1$ ($%
1/\alpha \omega \rightarrow \infty $), that provides an analytical solution
to the wave equation\ and as a second case, an arbitrary relation $\alpha
\omega $ can be considered.

\section{Spin-0 equation with vector and scalar potentials of the Coulomb
types in the space-time of a topological defect}

The Dirac equation is a wave equation that represents very well spin-1/2
particles in Minkowski space-time. The spin-0 particles are represented by
the usual Klein--Gordon equation which can be generalized to the curved
space-time case. In order to determine the generalization of the wave
equation one may replace the ordinary derivatives by covariant derivative 
\cite{santos2} in the spin-0 equation in Minkowski space-time, the result is 
\begin{equation}
-\frac{1}{\sqrt{-g}}D_{\mu }\left( g^{\mu \nu }\sqrt{-g}D_{\nu }\psi \right)
+m^{2}\psi =0,  \label{eq5}
\end{equation}%
that is the Klein--Gordon equation in a curved space-time \cite{birrel},
where $m$ is the particle mass, $D_{\mu }=\partial _{\mu }-ieA_{\mu }$, and $%
e$ is the electric charge. A scalar potential $V\left( r\right) $ may be
taken into account by making a modification on the mass term: $m\rightarrow
m+V\left( r\right) $. Substituting this mass term into (\ref{eq5}) we obtain
the following differential equation%
\begin{equation}
-\frac{1}{\sqrt{-g}}D_{\mu }\left( g^{\mu \nu }\sqrt{-g}D_{\nu }\psi \right)
+\left( m+V\right) ^{2}\psi =0.  \label{eq5b}
\end{equation}%
This differential equation takes into account a scalar potential $V$ and a
potential vector $A_{\mu }$ \cite{string12,string13}$.$ In the following, we
will obtain two classes of solutions of equation $\left( \ref{eq5b}\right) $%
. First, we will consider a slow rotation regime, and then we consider an
arbitrary relation $\alpha \omega $ in the space-time.

By considering the line element (\ref{eq3}) and the potential vector $A_{0}$%
, we obtain the following differential equation

\begin{eqnarray}
&&\left[ -\left( \frac{\partial }{\partial t}-ieA_{0}\right) ^{2}+\frac{1}{r}%
\left( \frac{\partial }{\partial r}\right) r\left( \frac{\partial }{\partial
r}\right) \right.   \label{eq6} \\
&&+\frac{\partial ^{2}}{\partial z^{2}}+\left( \frac{1-\alpha
^{2}r^{2}\omega ^{2}}{\alpha ^{2}r^{2}}\right) \frac{\partial ^{2}}{\partial
\phi ^{2}} \\
&&\left. \left. +2\omega \left( \frac{\partial }{\partial \phi }\right)
\left( \frac{\partial }{\partial t}+ieA_{0}\right) -\left( m+V\right) ^{2}%
\right] \phi =0\right. 
\end{eqnarray}%
that is the spin-0 equation in the space-time of topological defects. One
can see that equation (\ref{eq6}) is independent of $t$, $z$ and $\phi $, so
it is reasonable to write the solution as 
\begin{equation}
\psi \left( t,r,z,\phi \right) =e^{-i\varepsilon t}e^{il\phi
}e^{ip_{z}z}R\left( r\right) ,  \label{eq7}
\end{equation}%
where $l=0,\pm 1,\pm 2,\pm 3,...$, and $\varepsilon $ can be interpreted as
the energy of the particle, $p_{z}$ is the momentum. Substituting (\ref{eq7}%
) into Eq. (\ref{eq6}), and by considering $A_{0}=$ $\kappa /r,$ we obtain
the radial differential equation 
\begin{equation*}
\left[ \frac{d^{2}}{dr^{2}}+\frac{1}{r}\frac{d}{dr}+\frac{e^{2}\kappa
^{2}-l^{2}/\alpha ^{2}}{r^{2}}-\left( m+V\right) ^{2}\right. 
\end{equation*}%
\begin{equation}
\left. \frac{2\varepsilon e\kappa +2\omega e\kappa l}{r}+\left( \varepsilon
+\omega l\right) ^{2}-p_{z}^{2}\right] R\left( r\right) =0,  \label{eq8}
\end{equation}%
where the parameter $\alpha $ represents the deficit angle of the space-time
and $\alpha =1$ corresponds to the Minkowski space-time. In this paper, we
are interested in studying the case $\alpha <1.$

In this stage, we consider a scalar potential of the type $V\left( r\right)
=\eta /r$, where $\eta $ is a constant, so, substituting this potential into
Eq. (\ref{eq8}), we get \ 
\begin{equation}
\left[ \frac{d^{2}}{dr^{2}}+\frac{1}{r}\frac{d}{dr}-\frac{\beta ^{2}}{r^{2}}-%
\frac{2\gamma }{r}-\delta ^{2}\right] R\left( r\right) =0,  \label{eq9}
\end{equation}%
where 
\begin{equation}
\delta ^{2}=m^{2}+p_{z}^{2}-\left( \varepsilon +\omega l\right) ^{2},\text{
\ }\beta =\sqrt{l^{2}/\alpha ^{2}+\eta ^{2}-e^{2}\kappa ^{2}},\text{ \ }%
\gamma =-\varepsilon e\kappa -\omega e\kappa l+m\eta .  \label{eq9b}
\end{equation}
We assume the relation $e^2\kappa^2 < \eta^2$ so that $\beta$ is a real
number. Now, we will consider a transformation of the radial coordinate 
\begin{equation}
\rho =2\delta r,  \label{eq10}
\end{equation}%
and as a result, Eq.$\left( \ref{eq9}\right) $ will take the form%
\begin{equation}
\left[ \frac{d^{2}}{d\rho ^{2}}+\frac{1}{\rho }\frac{dR}{d\rho }-\frac{\beta
^{2}}{\rho ^{2}}-\frac{\gamma }{\delta \rho }-\frac{1}{4}\right] R\left(
\rho \right) =0.  \label{eq11}
\end{equation}%
Normalizable eigenfunctions may be obtained if we propose the solution%
\begin{equation}
R\left( \rho \right) =\rho ^{\beta }e^{-\frac{\rho }{2}}F\left( \rho \right)
,  \label{eq12}
\end{equation}%
then substituting $R(\rho )$ (\ref{eq12}) into eq. (\ref{eq11}), \ we obtain
the differential equation that can be associated with the radial equation%
\begin{equation}
\rho \frac{d^{2}F}{d\rho ^{2}}+\left( 2\beta +1-\rho \right) \frac{dF}{d\rho 
}+\left( -\beta -\frac{\gamma }{\delta }-\frac{1}{2}\right) F=0.
\label{eq13}
\end{equation}%
This is the confluent hypergeometric equation, which is a second order homogeneous differential equation where two independent solutions
can be obtained. The solution of Eq. (\ref{eq13}), regular at $\rho =0,$ is
given by the confluent hypergeometric function that is denoted by 
\begin{equation}
F\left( \rho \right) =_{1}F_{1}\left( \beta +\frac{\gamma }{\delta }+\frac{1%
}{2},2\beta +1;\rho \right) .  \label{eq14}
\end{equation}

\begin{figure}[b]
\includegraphics[scale=0.7]{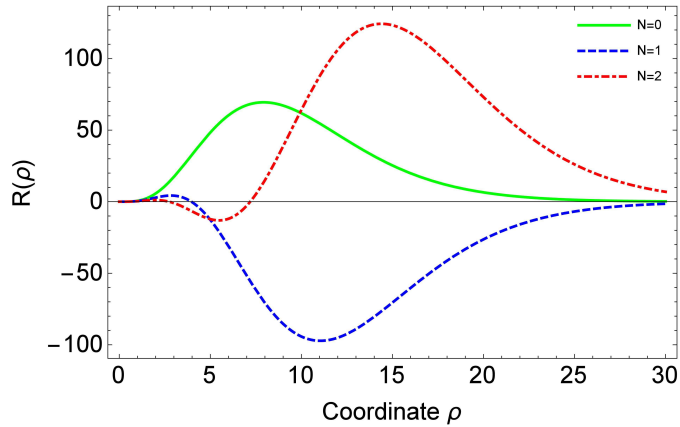}\newline
\caption{The plots of the radial coordinate $R$\ as the function of the
variable $\protect\rho$\ displayed for three different values of $N$ with
the parameters $\protect\alpha=0.9$, $\protect\omega=0.6$, $\protect\eta=1$
and $l=1.$}
\label{fig1}
\end{figure}

If we consider the limit $\alpha \omega \ll 1$, that is a slow rotation
regime, the boundary condition implies that the solution%
\begin{equation}
_{1}F_{1}\left( \beta +\frac{\gamma }{\delta }+\frac{1}{2},2\beta +1;\frac{%
2\delta }{\alpha \omega }\rightarrow \infty \right)  \label{eq15}
\end{equation}%
must be a finite as $\rho _{0}=1/\alpha \omega \rightarrow \infty $. So, due
to asymptotic behavior of the hypergeometric function, it is necessary that
the $_{1}F_{1}$ function be a polynomial function of degree $N$ and the
parameter $\beta +\frac{\gamma }{\delta }+\frac{1}{2}$ should be a negative
integer. These conditions implies that 
\begin{equation}
\beta +\frac{\gamma }{\delta }+\frac{1}{2}=-N,\text{ \ \ \ }N=0,1,2,...\text{
,}  \label{eq16}
\end{equation}%
and combining this equation and equation $\left( \ref{eq9b}\right) $ we
finally obtain the spectrum of energy%
\begin{equation}
\varepsilon =\frac{e\kappa m\eta }{\zeta +e^{2}\kappa ^{2}}\pm \sqrt{\zeta 
\frac{\left[ \left( p_{z}^{2}+m^{2}\right) \left( \zeta +e^{2}\kappa
^{2}\right) -m^{2}\eta ^{2}\right] }{\left( \zeta +e^{2}\kappa ^{2}\right)
^{2}}}-\omega \left\vert l\right\vert ,  \label{eq17}
\end{equation}%
where $\zeta =\left( N+\frac{1}{2}+\sqrt{l^{2}/\alpha ^{2}+\eta
^{2}-e^{2}\kappa ^{2}}\right) ^{2}$.

\begin{figure}[h]
\includegraphics[scale=0.7]{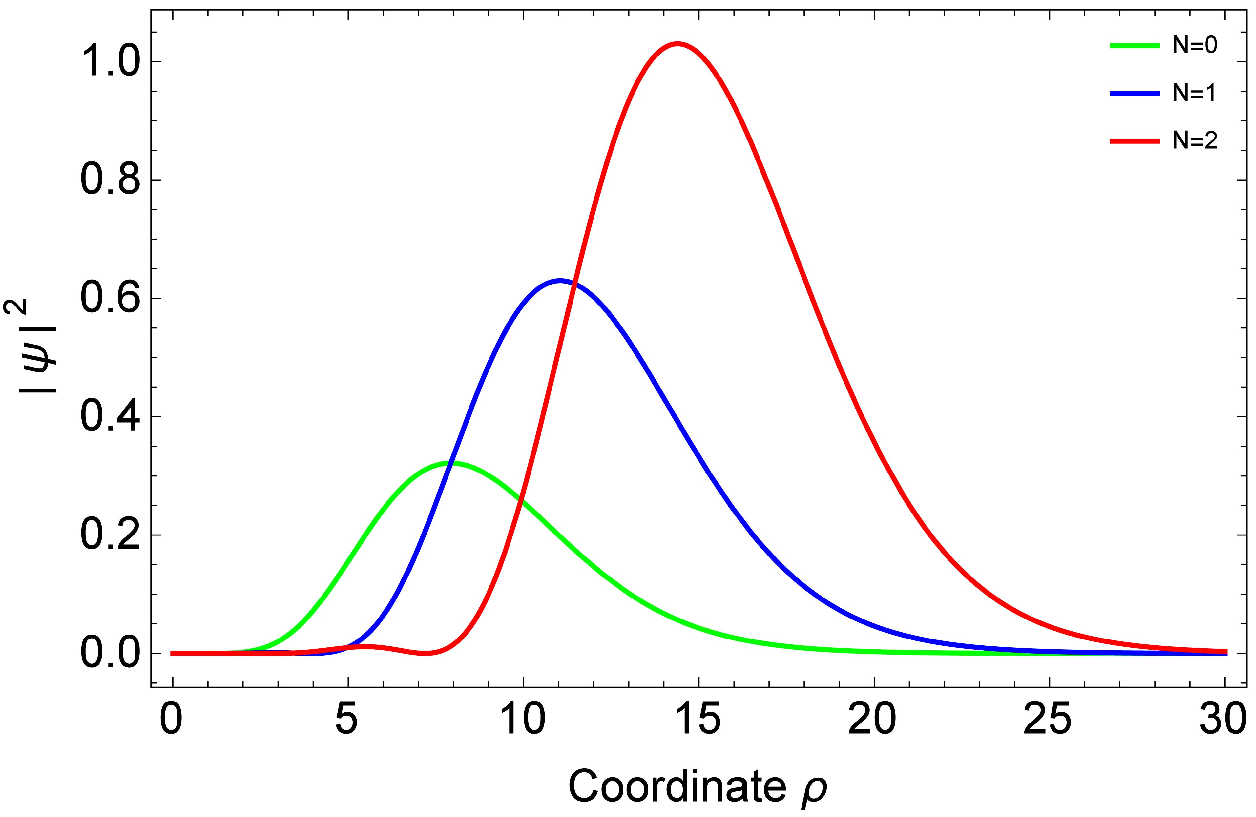}\newline
\caption{The plots of $\left\vert \protect\psi\right\vert ^{2}$\ \ as the
function of the variable $\protect\rho$ displayed for three different values
of $N$\ with the parameters $\protect\alpha=0.9$, $\protect\omega=0.6$, $%
\protect\eta=1$\ and $l=1.$}
\end{figure}

Observing Eq. $\left( \ref{eq17}\right) $, we can see that the energy
spectrum depends on $\alpha $, the deficit angle of the conical space-time.
The first and second terms are associated to the Coulomb-like potentials
embedded in a cosmic string background and the third term is associated to
the non-inertial effect of rotational frames, that is a Page-Werner et al. term 
\cite{inercial5,inercial1,inercial2,inercial3}. For $l=0$ or $\omega =0$ the
discrete set of energies are symmetrical about $\varepsilon =0$, in this
way, the presence of non-inertial effects of rotational frames in space-time
breaks the symmetry of energy levels about $\varepsilon =0$ because $%
\varepsilon _{+}$, in general, is greater than $\varepsilon _{-}$.

\begin{figure}[h]
\includegraphics[scale=0.4]{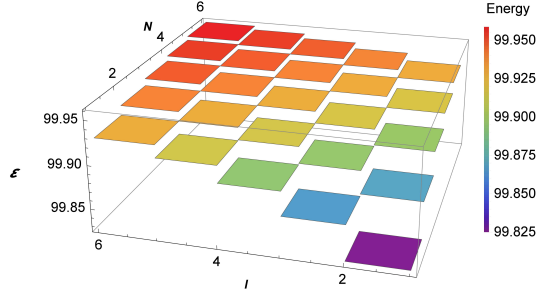}\newline
\caption{The plots of particle energy spectrum $\protect\varepsilon$ as the
function of variables $N$ and $l$. }
\end{figure}

From equation $\left( \ref{eq17}\right) $, it is possible to see that the
energy depends on the constant $\alpha $, thus the presence of the
topological defect modifies the energy of the particle.

\begin{figure}[h]
\includegraphics[scale=0.4]{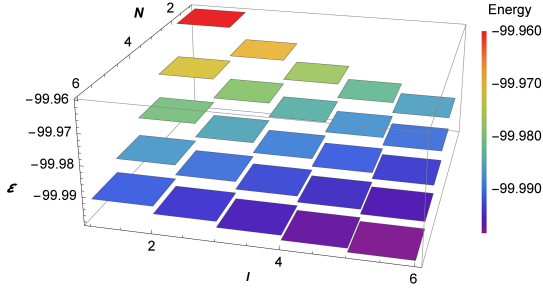}\newline
\caption{The plots of negative energy spectrum $\protect\varepsilon$ as the
function of variables $N$\ and $l$.}
\end{figure}

Fig. 1 and 2 show that the radial solution $R\left( \rho \right) $ decreases
with the coordinate $\rho $ and becomes negligible far away from the
topological defect as $\rho \rightarrow \infty $. For clarity, the plots of
the energy spectrum $\varepsilon $ as function of the variables $N$ and $l$
are shown in Figures 5 and 6.

In the next section we will discuss an arbitrary relation $\omega\alpha$ for
the Klein--Gordon oscillator where the shape of the potential is adequate for
this purpose.

\section{Klein--Gordon oscillator in the the space-time of topological defects%
}

Another system of interest that may be considered is the Klein--Gordon
oscillator \cite{kgoscillator1} in the background of the cosmic string
space-time. In recent years, several studies have addressed the Klein--Gordon
oscillator in quantum systems \cite%
{kgoscillator2,kgoscillator3,kgoscillator4,kgoscillator5,kgoscillator6,kgoscillator7,kgoscillator8,kgoscillator9,kgoscillator10}%
. It has a formulation similar to the vector potential in the previous
section, so to study its solutions we will use the following change in
momentum operator:%
\begin{equation}
p_{\mu }\rightarrow \left( p_{\mu }+im\Omega X_{\mu }\right) ,  \label{eq21}
\end{equation}%
where $m$ is the particle mass at rest, $\Omega $ is the frequency of the
oscillator and $X_{\mu }=\left( 0,r,0,0\right) $, with $r$ being the
distance from the particle to the string. In this way, the wave equation
becomes%
\begin{equation}
\left[ -\frac{1}{\sqrt{-g}}\left( \partial _{\mu }+m\Omega X_{\mu }\right)
g^{\mu \nu }\sqrt{-g}\left( \partial _{\nu }-m\Omega X_{\nu }\right) +\left(
m+V\right) ^{2}\right] \psi =0.  \label{eq22}
\end{equation}%
Taking $V=0$ in above equation and by considering the line element (\ref{eq3}%
), we obtain the following equation%
\begin{eqnarray}
&&\left[ -\frac{\partial ^{2}}{\partial t^{2}}+\frac{1}{r}\left( \frac{%
\partial }{\partial r}+m\Omega \right) r\left( \frac{\partial }{\partial r}%
-m\Omega \right) +\frac{\partial ^{2}}{\partial z^{2}}\right.   \notag \\
&&\left. \left. +\left( \frac{1-\alpha ^{2}r^{2}\omega ^{2}}{\alpha ^{2}r^{2}%
}\right) \frac{\partial ^{2}}{\partial \phi ^{2}}+2\omega \frac{\partial ^{2}%
}{\partial t\partial \phi }-m^{2}\right] \psi =0\right. .  \label{eq23}
\end{eqnarray}%
Similar to the case of the Coulomb potential in last section, the eq. $%
\left( \ref{eq23}\right) $ is independent of $t$, $z$ and $\phi $, so it is
appropriate to choose the ansatz%
\begin{equation}
\psi \left( t,r,z,\phi \right) =e^{-i\varepsilon t}e^{il\phi
}e^{ip_{z}z}R^{\prime }\left( r\right) ,  \label{eq24}
\end{equation}%
with $l=0,\pm 1,\pm 2,\pm 3$, and $\varepsilon $ being the energy of the
particle. Substituting (\ref{eq24}) into Eq. (\ref{eq23}), we obtain the
radial differential equation%
\begin{eqnarray}
&&\left[ +\frac{1}{r}\left( \frac{\partial }{\partial r}+m\Omega \right)
r\left( \frac{\partial }{\partial r}-m\Omega \right) +\right.   \notag \\
&&\left. \left. -\frac{l^{2}}{\alpha ^{2}r^{2}}+\left( \varepsilon +\omega
l\right) ^{2}-p_{z}^{2}-m^{2}\right] R^{\prime }\left( r\right) =0\right. 
\label{eq25}
\end{eqnarray}%
At this stage, we can consider the substitution $R^{\prime }\left( r\right)
=R\left( r\right) /\sqrt{r}$ in the equation $\left( \ref{eq25}\right) $ the
result is 
\begin{equation}
\left[ \frac{d^{2}}{dr^{2}}-m^{2}\Omega ^{2}r^{2}-\frac{\left( l^{2}/\alpha
^{2}-1/4\right) }{r^{2}}+K^{2}\right] R\left( r\right) =0,  \label{eq26}
\end{equation}%
with $K=\sqrt{\left( \varepsilon +\omega l\right)
^{2}-p_{z}^{2}-m^{2}-2m\Omega }$. That is the radial equation that describes
the Klein--Gordon oscillator in the space-time of a topological defect. In
order to obtain the solution to the above differential equation it is
necessary to analyze its asymptotic behavior for $r\rightarrow 0$ and $%
r\rightarrow r_{0}$ where $r_{0}\equiv 1/\omega \alpha $. In this way, a
regular solution at the origin is obtained if the solution of equation $%
\left( \ref{eq26}\right) $ has the form%
\begin{equation}
R\left( r\right) =r^{\left\vert \frac{l}{\alpha }\right\vert +\frac{1}{2}%
}e^{-m\Omega r^{2}/2}F\left( r\right) \text{.}  \label{eq27}
\end{equation}%
Substituting the above expression in Eq. $\left( \ref{eq26}\right) $ and by
introducing the following new variable $\rho =m\Omega r^{2}$, we can rewrite
the radial Eq. $\left( \ref{eq26}\right) $ in the form 
\begin{equation}
\rho \frac{d^{2}F\left( \rho \right) }{d\rho ^{2}}+\left( \left\vert \frac{l%
}{\alpha }\right\vert +1-\rho \right) \frac{dF\left( \rho \right) }{d\rho }%
-\left( \frac{l}{2\alpha }+\frac{1}{2}-\frac{K^{2}}{4m\Omega }\right)
F\left( \rho \right) =0.  \label{eq28}
\end{equation}%
The solution of Eq. (\ref{eq28}) is given by the confluent hypergeometric
function that is denoted by 
\begin{equation}
F\left( \rho \right) =_{1}F_{1}\left( A,B;\rho \right) \text{,}  \label{eq29}
\end{equation}%
where the parameters $A$, $B$ and $\rho $ are given by 
\begin{eqnarray}
A &=&\frac{1}{2}\left\vert \frac{l}{\alpha }\right\vert +\frac{1}{2}-\frac{%
K^{2}}{4m\Omega }\text{, }  \label{eq30} \\
B &=&\left\vert \frac{l}{\alpha }\right\vert +1\text{,}  \label{eq31} \\
\rho  &=&m\Omega r^{2}\text{.}  \label{eq32}
\end{eqnarray}

\begin{figure}[b]
\includegraphics[scale=0.7]{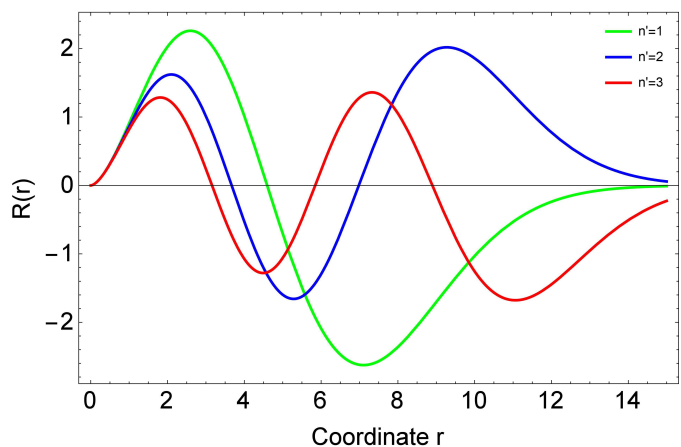}\newline
\caption{The plots of radial coordinate $R$ as the function of variable $r$
displayed for three different $n^{\prime }$ with the parameters $\protect%
\alpha=0.9$, $\protect\omega=0.6$, $\Omega=0.1$, $m=1$ and $l=1$.}
\label{fig5b}
\end{figure}

\subsection{Limit $\protect\alpha \protect\omega \ll 1$ ($1/\protect\alpha 
\protect\omega \rightarrow \infty $)}

Following the discussions of the Sect. 3, we proceed now to find the
eigenfunction for this problem.

\begin{figure}[h]
\includegraphics[scale=0.7]{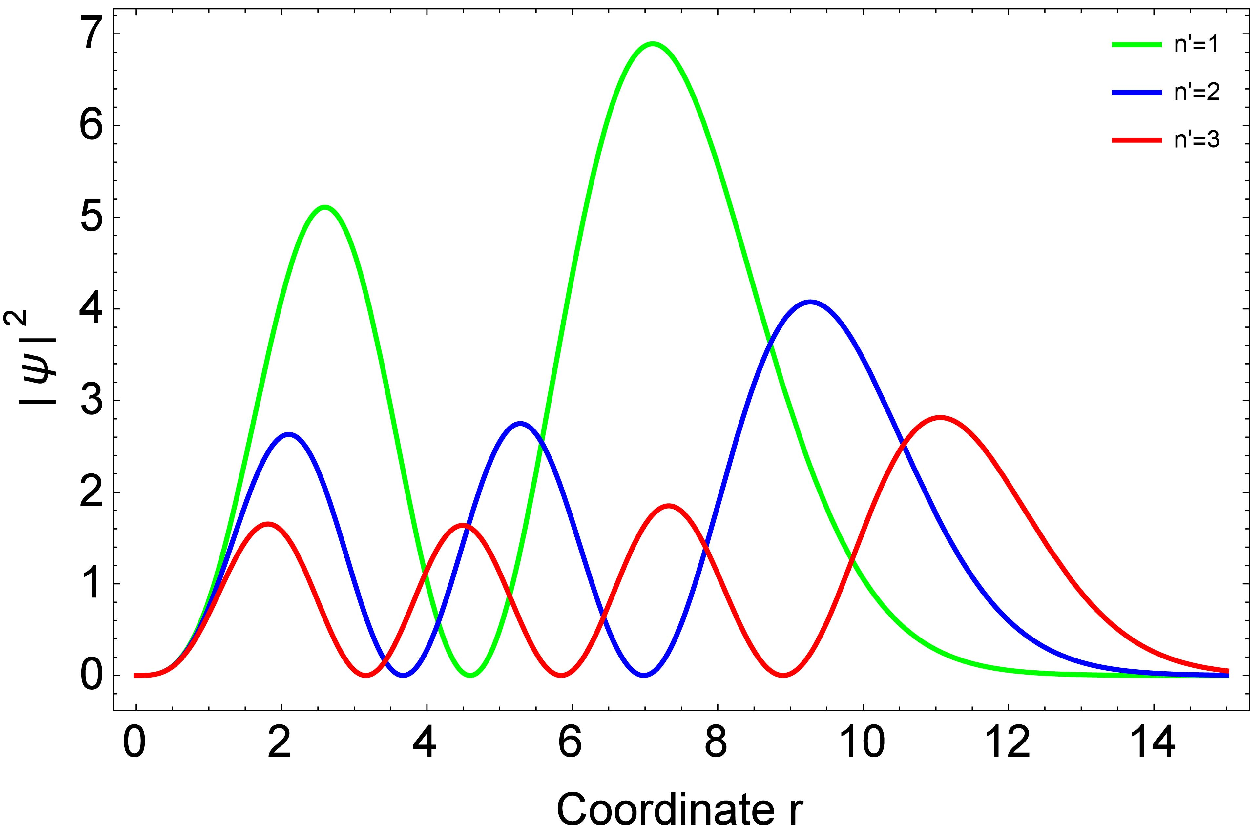}\newline
\caption{The plots of $\left\vert \protect\psi\right\vert ^{2}$\ \ as
functions of the variable $\protect\rho$ displayed for three different
values of $n^{\prime }$ with the parameters $\protect\alpha=0.9$, $\protect%
\omega=0.6$, $\Omega=0.1$, $m=1$ and $l=1.$}
\end{figure}

Considering again the limit $\alpha \omega \ll 1$, we have a change in the
boundary condition on the radial coordinate, i.e, when $\alpha \omega \ll 1$
the radial coordinate tends to infinity at $r=1/\omega \alpha $.
Consequently the hypergeometric function must be a polynomial function of
degree $N,$\ and the parameter $A=\frac{1}{2}\left\vert \frac{l}{\alpha }%
\right\vert +\frac{1}{2}-\frac{K^{2}}{4m\Omega }$, must be a negative
integer. This condition implies that%
\begin{equation}
\frac{1}{2}\left\vert \frac{l}{\alpha }\right\vert +\frac{1}{2}-\frac{K^{2}}{%
4m\Omega }=-N,  \label{eq35}
\end{equation}%
and by the use of the definition of 
\begin{equation}
K=\sqrt{\left( \varepsilon +\omega l\right) ^2-p_{z}^{2}-m^{2}-2m\Omega },
\end{equation}%
we finally obtain the set of energies%
\begin{equation}
\varepsilon =\pm \sqrt{2m\Omega \left( 2n^{\prime }+\left\vert \frac{l}{%
\alpha }\right\vert \right) +m^{2}+p_{z}^{2}}-\omega \left\vert l\right\vert
,\text{ \ \ }n^{\prime }\equiv N+1=1,2,3,...\text{ .}  \label{eq36}
\end{equation}

We can see that the energy spectrum associated with the Klein--Gordon
Oscillator in the conical space-time depends on $\alpha $, that is the
deficit angle of the conical space-time. It increases the energy of the
system if $\alpha <1$. It is easy to see that for $l=0$ or $\omega =0$ the
energy is symmetrical about $\varepsilon =0$. In this way, the rotating
reference system breaks the symmetry of the energy about $\varepsilon =0$. The first
term in Eq. $\left( \ref{eq36}\right) $ is associated to the Klein--Gordon
Oscillator embedded in a conical space and the second one is associated with
the non-inertial effect, which in turn is a coupling between the angular
quantum number and the angular velocity of the rotational reference system. As it may be
seen in Fig. 5, the radial eigenfunction becomes negligible far away from
the string as $\rho \rightarrow \infty .$ Fig. 6 presents $\left\vert \psi
\right\vert ^{2}$ as function of the variable $\rho $ for three different
values of $n^{\prime }$. The energy spectrum as a function of the variables $%
n^{\prime }$ and $l$ are shown in the plots of Figures 7 and 8. We note that
the result obtained in eq. \ref{eq36} is similar to the one reported in \cite%
{Castro2} for scalar bosons described by the Duffin--Kemmer--Petiau (DKP)
formalism.

\begin{figure}[h]
\includegraphics[scale=0.4]{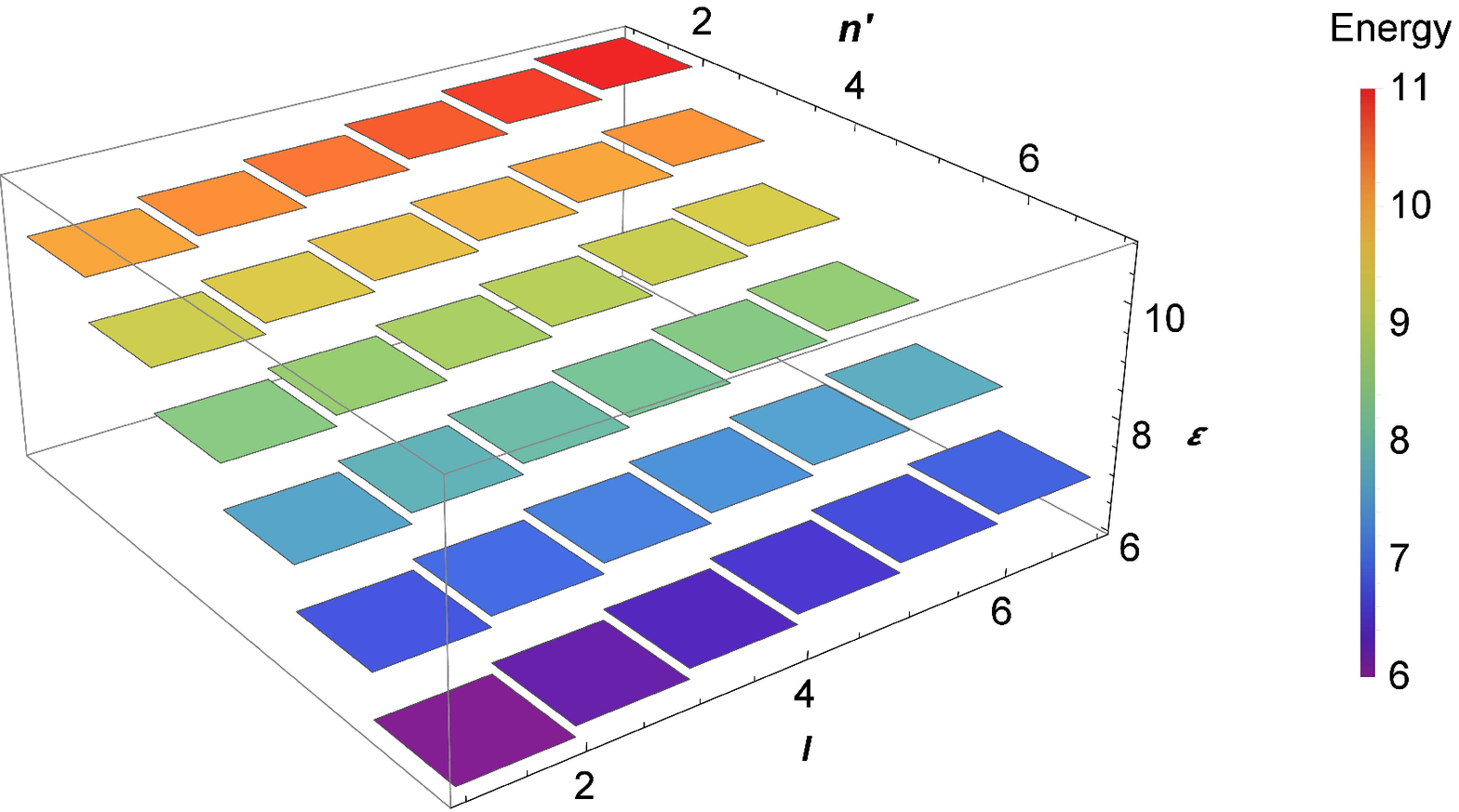}\newline
\caption{The plots the of particle energy spectrum $\protect\varepsilon$ as
function of the variables $n^{\prime }$ and $l$. }
\end{figure}

\begin{figure}[h]
\includegraphics[scale=0.4]{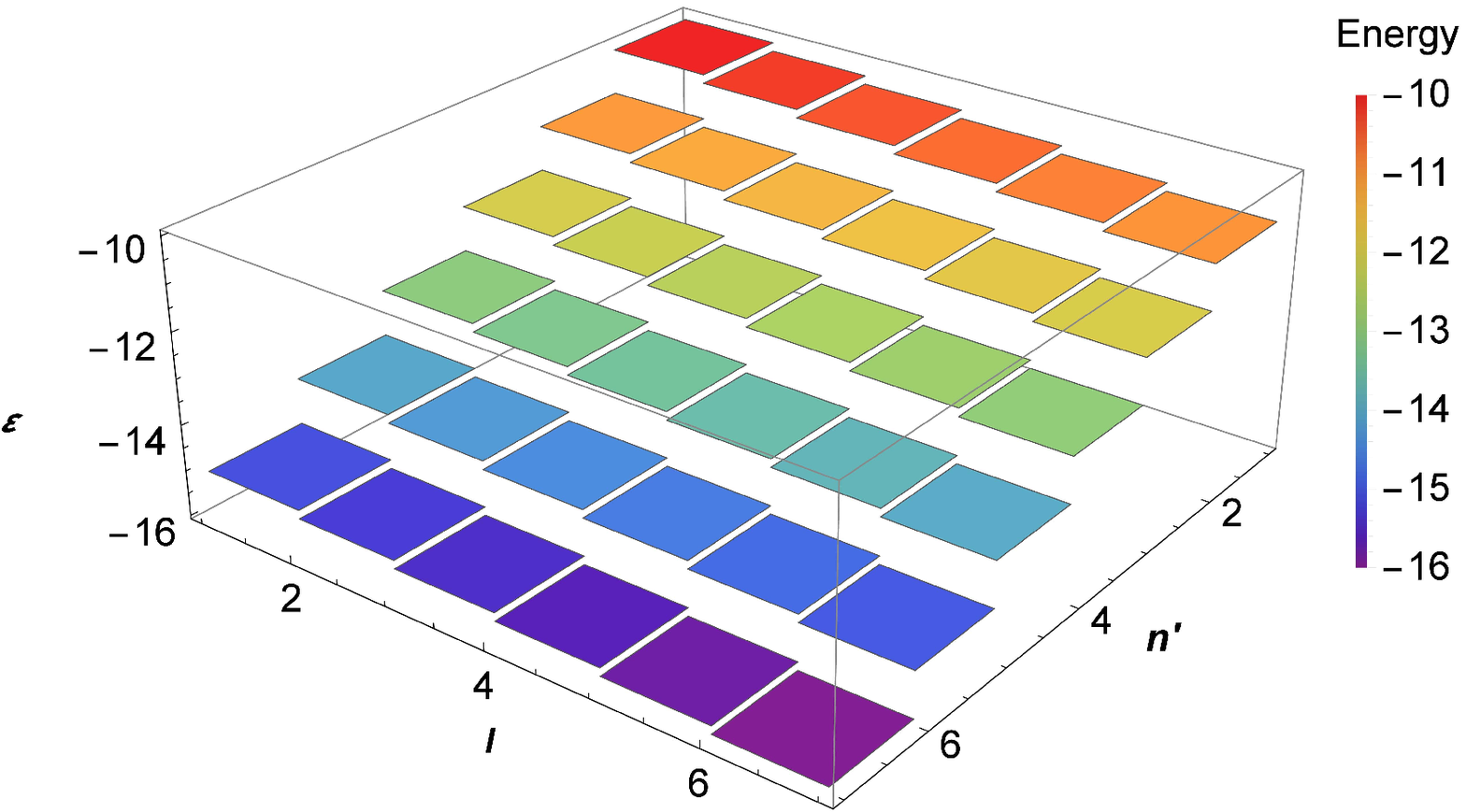}\newline
\caption{The plots of the negative energy spectrum $\protect\varepsilon$ as
function of the variables $n^{\prime }$\ and $l$.}
\end{figure}

\subsection{Arbitrary $\protect\omega \protect\alpha $}

Now let us study an arbitrary relation between the parameters $\alpha $ and $%
\omega .$ In this case we discuss the behaviour of the Klein--Gordon
oscillator without assuming the Limit $\alpha \omega \ll 1$ . The physical
condition implies that the wave function vanishes at $r_{0} =1/\alpha \omega 
$, i.e.,

\begin{equation}
_{1}F_{1}\left( A,B;\rho_{0}=m\Omega r_{0}^2\right) =0.  \label{eq37}
\end{equation}

If one assumes that $m\Omega \ll 1$, the parameter $A$ of the hypergeometric
function can be considered large and the parameters $B$ and $r_{0}$ remain
fixed. So, we can use these results to expand the hypergeometric function in
the form \cite{abramo,inercial10}

\begin{align}
_{1}F_{1}(A,B;\rho_{0}) &\approx\dfrac{\Gamma(B)}{\sqrt{\pi}}e^{\dfrac{%
\rho_{0}}{2}}\left(\dfrac{B\rho_{0}}{2}-A\rho_{0}\right) ^{\dfrac{1-B}{2}%
}\times  \notag \\
& \times\cos\left(\sqrt{2B\rho_{0}-4A\rho_{0}}-\dfrac{B\pi}{2}+\dfrac{\pi}{4}%
\right),  \label{eq38}
\end{align}
here $\Gamma(B)$ is the gamma function. By considering the condition $(\ref%
{eq37})$ and the eq. $(\ref{eq38})$ , we finally obtain the set of energies
for an arbitrary relation between the parameters $\alpha $ and $\omega $

\begin{equation}
\varepsilon \approx \sqrt{p_{z}^{2}+m^{2}+2m\Omega+\dfrac{1}{r_{0}^2}\left(%
\dfrac{l\pi}{2\alpha}+\dfrac{3\pi}{4}+n\pi\right)^{2}}-\omega l  \label{eq39}
\end{equation}
where $n=0,1,2,... $ is the radial quantum number of solution. Then equation 
$(\ref{eq39})$ corresponds to the set of energies for an arbitrary relation
between $\alpha $ and $\omega$ where the non-inertial effects play o hole of
an hard-wall confining potential \cite{inercial10}. The first term of eq. $(%
\ref{eq39})$ is associated to the Klein--Gordon Oscillator embedded in a
cosmic string background and the second term is associated to the
non-inertial effect of rotational frames, which in turn is a Sagnac-type effect.

\section{Conclusions}

In this work, we have determined the spin-0 equation in the presence of a
vector and a scalar potential and examined the wave equation in the presence
of a Klein--Gordon oscillator in a cosmic string space-time. Despite the
complexity of the studied systems, we obtained compact expressions for the
energy spectrum and for the particles wave functions . It has been shown
that the potentials studied allow the formation of bound states and the
energy spectrum associated with the relativistic wave equation in a cosmic
string space depends on the deficit angle $\alpha $, this fact shows that
the topological defect modifies the energy of physical systems.

An important result that we have shown is that the non-inertial effect
restrict the region of the space-time where the particle can be observed and
beyond that, it shifts the energy levels. This feature reveals the existence
of a coupling between the angular quantum number and the angular velocity of
the rotational reference system.

We have shown that the wave equation presents two different classes of
solutions that depend on the value of the $\alpha $ and $\omega $. In the
first case it is assumed the limit $\alpha \omega \ll 1$, that is a slow
rotation regime, and as a second case, it is considered an arbitrary
relation $\alpha \omega .$ For both classes of solutions, we have found the
energy spectrum and the eigenfunctions and we have shown that the discrete
set of energies in general is composed of two contributions. The first term
is associated to the external potential embedded in a cosmic string
background and the second one is associated to non-inertial effects. With
these results it is possible to have an idea about the general aspects of
the quantum dynamics of scalar bosons inside a cosmic string background.

So, in this paper, we have shown some results about quantum systems where
general relativistic effects are take into account, that in addition with
the previous results \cite{santos1,santos2} present many interesting
effects. This is a fundamental subject in physics, and the connections
between these theories are not well understood.

\bigskip

\section{Acknowledgments}

This work was supported in part by means of funds provided by CAPES.

\bibliographystyle{aipnum4-1}
\bibliography{referencias_unificadas}

\begin{thebibliography}{50}%
\makeatletter
\providecommand \@ifxundefined [1]{%
 \@ifx{#1\undefined}
}%
\providecommand \@ifnum [1]{%
 \ifnum #1\expandafter \@firstoftwo
 \else \expandafter \@secondoftwo
 \fi
}%
\providecommand \@ifx [1]{%
 \ifx #1\expandafter \@firstoftwo
 \else \expandafter \@secondoftwo
 \fi
}%
\providecommand \natexlab [1]{#1}%
\providecommand \enquote  [1]{``#1''}%
\providecommand \bibnamefont  [1]{#1}%
\providecommand \bibfnamefont [1]{#1}%
\providecommand \citenamefont [1]{#1}%
\providecommand \href@noop [0]{\@secondoftwo}%
\providecommand \href [0]{\begingroup \@sanitize@url \@href}%
\providecommand \@href[1]{\@@startlink{#1}\@@href}%
\providecommand \@@href[1]{\endgroup#1\@@endlink}%
\providecommand \@sanitize@url [0]{\catcode `\\12\catcode `\$12\catcode
  `\&12\catcode `\#12\catcode `\^12\catcode `\_12\catcode `\%12\relax}%
\providecommand \@@startlink[1]{}%
\providecommand \@@endlink[0]{}%
\providecommand \url  [0]{\begingroup\@sanitize@url \@url }%
\providecommand \@url [1]{\endgroup\@href {#1}{\urlprefix }}%
\providecommand \urlprefix  [0]{URL }%
\providecommand \Eprint [0]{\href }%
\providecommand \doibase [0]{http://dx.doi.org/}%
\providecommand \selectlanguage [0]{\@gobble}%
\providecommand \bibinfo  [0]{\@secondoftwo}%
\providecommand \bibfield  [0]{\@secondoftwo}%
\providecommand \translation [1]{[#1]}%
\providecommand \BibitemOpen [0]{}%
\providecommand \bibitemStop [0]{}%
\providecommand \bibitemNoStop [0]{.\EOS\space}%
\providecommand \EOS [0]{\spacefactor3000\relax}%
\providecommand \BibitemShut  [1]{\csname bibitem#1\endcsname}%
\let\auto@bib@innerbib\@empty
\bibitem [{\citenamefont {Castro}(2016)}]{Castro1}%
  \BibitemOpen
  \bibfield  {author} {\bibinfo {author} {\bibfnamefont {L.~B.}\ \bibnamefont
  {Castro}},\ }\href {\doibase 10.1140/epjc/s10052-016-3904-4} {\bibfield
  {journal} {\bibinfo  {journal} {Eur. Phys. J. C}\ }\textbf {\bibinfo {volume}
  {76}},\ \bibinfo {pages} {1} (\bibinfo {year} {2016})}\BibitemShut {NoStop}%
\bibitem [{\citenamefont {Castro}(2015)}]{Castro2}%
  \BibitemOpen
  \bibfield  {author} {\bibinfo {author} {\bibfnamefont {L.~B.}\ \bibnamefont
  {Castro}},\ }\href {\doibase 10.1140/epjc/s10052-015-3507-5} {\bibfield
  {journal} {\bibinfo  {journal} {Eur. Phys. J. C}\ }\textbf {\bibinfo {volume}
  {75}},\ \bibinfo {pages} {1} (\bibinfo {year} {2015})}\BibitemShut {NoStop}%
\bibitem [{\citenamefont {Filho}\ \emph {et~al.}(2016)\citenamefont {Filho},
  \citenamefont {Braga}, \citenamefont {Lira},\ and\ \citenamefont
  {Jr.}}]{incert1}%
  \BibitemOpen
  \bibfield  {author} {\bibinfo {author} {\bibfnamefont {R.~N.~C.}\
  \bibnamefont {Filho}}, \bibinfo {author} {\bibfnamefont {J.~P.}\ \bibnamefont
  {Braga}}, \bibinfo {author} {\bibfnamefont {J.~H.}\ \bibnamefont {Lira}}, \
  and\ \bibinfo {author} {\bibfnamefont {J.~S.~A.}\ \bibnamefont {Jr.}},\
  }\href {\doibase http://dx.doi.org/10.1016/j.physletb.2016.02.035} {\bibfield
   {journal} {\bibinfo  {journal} {Phys. Lett. A}\ }\textbf {\bibinfo {volume}
  {755}},\ \bibinfo {pages} {367 } (\bibinfo {year} {2016})}\BibitemShut
  {NoStop}%
\bibitem [{\citenamefont {Parker}(1980)}]{parker1}%
  \BibitemOpen
  \bibfield  {author} {\bibinfo {author} {\bibfnamefont {L.}~\bibnamefont
  {Parker}},\ }\href@noop {} {\bibfield  {journal} {\bibinfo  {journal} {Phys.
  Rev. Lett.}\ }\textbf {\bibinfo {volume} {44}},\ \bibinfo {pages} {1559}
  (\bibinfo {year} {1980})}\BibitemShut {NoStop}%
\bibitem [{\citenamefont {Marques}\ and\ \citenamefont
  {Bezerra}(2002)}]{hcurvo1}%
  \BibitemOpen
  \bibfield  {author} {\bibinfo {author} {\bibfnamefont {G.}~\bibnamefont
  {Marques}}\ and\ \bibinfo {author} {\bibfnamefont {V.}~\bibnamefont
  {Bezerra}},\ }\href
  {http://www.scopus.com/inward/record.url?eid=2-s2.0-0242267817&partnerID=40&md5=f9f8c34334692b6247486472e2bd802d}
  {\bibfield  {journal} {\bibinfo  {journal} {Phys. Rev. D}\ }\textbf {\bibinfo
  {volume} {66}},\ \bibinfo {pages} {105011} (\bibinfo {year}
  {2002})}\BibitemShut {NoStop}%
\bibitem [{\citenamefont {Barros}(2005)}]{Barros1}%
  \BibitemOpen
  \bibfield  {author} {\bibinfo {author} {\bibfnamefont {C.~C.}\ \bibnamefont
  {Barros}},\ }\href@noop {} {\bibfield  {journal} {\bibinfo  {journal} {Eur.
  Phys. J. C}\ }\textbf {\bibinfo {volume} {42}},\ \bibinfo {pages} {119}
  (\bibinfo {year} {2005})}\BibitemShut {NoStop}%
\bibitem [{\citenamefont {Barros}(2006)}]{barros3}%
  \BibitemOpen
  \bibfield  {author} {\bibinfo {author} {\bibfnamefont {C.}~\bibnamefont
  {Barros}},\ }\href@noop {} {\bibfield  {journal} {\bibinfo  {journal} {Eur.
  Phys. J. C}\ }\textbf {\bibinfo {volume} {45}},\ \bibinfo {pages} {421}
  (\bibinfo {year} {2006})}\BibitemShut {NoStop}%
\bibitem [{\citenamefont {Carvalho}, \citenamefont {de~M.~Carvalho},\ and\
  \citenamefont {Furtado}(2014)}]{string8}%
  \BibitemOpen
  \bibfield  {author} {\bibinfo {author} {\bibfnamefont {J.}~\bibnamefont
  {Carvalho}}, \bibinfo {author} {\bibfnamefont {A.}~\bibnamefont
  {de~M.~Carvalho}}, \ and\ \bibinfo {author} {\bibfnamefont {C.}~\bibnamefont
  {Furtado}},\ }\href {\doibase 10.1140/epjc/s10052-014-2935-y} {\bibfield
  {journal} {\bibinfo  {journal} {Eur. Phys. J. C}\ }\textbf {\bibinfo {volume}
  {74}} (\bibinfo {year} {2014}),\ 10.1140/epjc/s10052-014-2935-y}\BibitemShut
  {NoStop}%
\bibitem [{\citenamefont {Brill}\ and\ \citenamefont
  {Wheeler}(1957)}]{neutrino1}%
  \BibitemOpen
  \bibfield  {author} {\bibinfo {author} {\bibfnamefont {D.~R.}\ \bibnamefont
  {Brill}}\ and\ \bibinfo {author} {\bibfnamefont {J.~A.}\ \bibnamefont
  {Wheeler}},\ }\href {\doibase 10.1103/RevModPhys.29.465} {\bibfield
  {journal} {\bibinfo  {journal} {Rev.Mod.Phys.}\ }\textbf {\bibinfo {volume}
  {29}},\ \bibinfo {pages} {465} (\bibinfo {year} {1957})}\BibitemShut
  {NoStop}%
\bibitem [{\citenamefont {Linde}(1979)}]{string1}%
  \BibitemOpen
  \bibfield  {author} {\bibinfo {author} {\bibfnamefont {A.~D.}\ \bibnamefont
  {Linde}},\ }\href {http://stacks.iop.org/0034-4885/42/i=3/a=001} {\bibfield
  {journal} {\bibinfo  {journal} {Reports on Progress in Physics}\ }\textbf
  {\bibinfo {volume} {42}},\ \bibinfo {pages} {389} (\bibinfo {year}
  {1979})}\BibitemShut {NoStop}%
\bibitem [{\citenamefont {Vilenkin}(1981)}]{string2}%
  \BibitemOpen
  \bibfield  {author} {\bibinfo {author} {\bibfnamefont {A.}~\bibnamefont
  {Vilenkin}},\ }\href {\doibase 10.1103/PhysRevD.24.2082} {\bibfield
  {journal} {\bibinfo  {journal} {Phys. Rev. D}\ }\textbf {\bibinfo {volume}
  {24}},\ \bibinfo {pages} {2082} (\bibinfo {year} {1981})}\BibitemShut
  {NoStop}%
\bibitem [{\citenamefont {Germano}, \citenamefont {Bezerra},\ and\
  \citenamefont {de~Mello}(1996)}]{string7}%
  \BibitemOpen
  \bibfield  {author} {\bibinfo {author} {\bibfnamefont {M.~G.}\ \bibnamefont
  {Germano}}, \bibinfo {author} {\bibfnamefont {V.~B.}\ \bibnamefont
  {Bezerra}}, \ and\ \bibinfo {author} {\bibfnamefont {E.~R.~B.}\ \bibnamefont
  {de~Mello}},\ }\href {http://stacks.iop.org/0264-9381/13/i=10/a=006}
  {\bibfield  {journal} {\bibinfo  {journal} {Class. Quant. Grav.}\ }\textbf
  {\bibinfo {volume} {13}},\ \bibinfo {pages} {2663} (\bibinfo {year}
  {1996})}\BibitemShut {NoStop}%
\bibitem [{\citenamefont {{\"O}lmez}, \citenamefont {Mandic},\ and\
  \citenamefont {Siemens}(2010)}]{string11}%
  \BibitemOpen
  \bibfield  {author} {\bibinfo {author} {\bibfnamefont {S.}~\bibnamefont
  {{\"O}lmez}}, \bibinfo {author} {\bibfnamefont {V.}~\bibnamefont {Mandic}}, \
  and\ \bibinfo {author} {\bibfnamefont {X.}~\bibnamefont {Siemens}},\
  }\href@noop {} {\bibfield  {journal} {\bibinfo  {journal} {Phys. Rev. D}\
  }\textbf {\bibinfo {volume} {81}},\ \bibinfo {pages} {104028} (\bibinfo
  {year} {2010})}\BibitemShut {NoStop}%
\bibitem [{\citenamefont {Abbott}\ \emph {et~al.}(2016)\citenamefont {Abbott},
  \citenamefont {Abbott}, \citenamefont {Abbott}, \citenamefont {Zhang},
  \citenamefont {Zhao}, \citenamefont {Zhou}, \citenamefont {Zhou},
  \citenamefont {Zhu}, \citenamefont {Zucker}, \citenamefont {Zuraw},\ and\
  \citenamefont {Zweizig}}]{waves1}%
  \BibitemOpen
  \bibfield  {author} {\bibinfo {author} {\bibfnamefont {B.~P.}\ \bibnamefont
  {Abbott}}, \bibinfo {author} {\bibfnamefont {R.}~\bibnamefont {Abbott}},
  \bibinfo {author} {\bibfnamefont {M.}~\bibnamefont {Abbott}}, \bibinfo
  {author} {\bibfnamefont {Y.}~\bibnamefont {Zhang}}, \bibinfo {author}
  {\bibfnamefont {C.}~\bibnamefont {Zhao}}, \bibinfo {author} {\bibfnamefont
  {M.}~\bibnamefont {Zhou}}, \bibinfo {author} {\bibfnamefont {Z.}~\bibnamefont
  {Zhou}}, \bibinfo {author} {\bibfnamefont {X.~J.}\ \bibnamefont {Zhu}},
  \bibinfo {author} {\bibfnamefont {M.~E.}\ \bibnamefont {Zucker}}, \bibinfo
  {author} {\bibfnamefont {S.~E.}\ \bibnamefont {Zuraw}}, \ and\ \bibinfo
  {author} {\bibfnamefont {J.}~\bibnamefont {Zweizig}} (\bibinfo
  {collaboration} {LIGO Scientific Collaboration and Virgo Collaboration}),\
  }\href {\doibase 10.1103/PhysRevLett.116.061102} {\bibfield  {journal}
  {\bibinfo  {journal} {Phys. Rev. Lett.}\ }\textbf {\bibinfo {volume} {116}},\
  \bibinfo {pages} {061102} (\bibinfo {year} {2016})}\BibitemShut {NoStop}%
\bibitem [{\citenamefont {Santos}\ and\ \citenamefont
  {Barros}(2016)}]{santos1}%
  \BibitemOpen
  \bibfield  {author} {\bibinfo {author} {\bibfnamefont {L.~C.~N.}\
  \bibnamefont {Santos}}\ and\ \bibinfo {author} {\bibfnamefont {C.~C.}\
  \bibnamefont {Barros}},\ }\href@noop {} {\bibfield  {journal} {\bibinfo
  {journal} {Eur. Phys. J. C}\ }\textbf {\bibinfo {volume} {76}},\ \bibinfo
  {pages} {560} (\bibinfo {year} {2016})}\BibitemShut {NoStop}%
\bibitem [{\citenamefont {Santos}\ and\ \citenamefont
  {Barros}(2017{\natexlab{a}})}]{santos3}%
  \BibitemOpen
  \bibfield  {author} {\bibinfo {author} {\bibfnamefont {L.~C.~N.}\
  \bibnamefont {Santos}}\ and\ \bibinfo {author} {\bibfnamefont {C.~C.}\
  \bibnamefont {Barros}},\ }\href@noop {} {\bibfield  {journal} {\bibinfo
  {journal} {arXiv preprint arXiv:1704.00408}\ } (\bibinfo {year}
  {2017}{\natexlab{a}})}\BibitemShut {NoStop}%
\bibitem [{\citenamefont {Gr{\o}n}(1975)}]{inercial12}%
  \BibitemOpen
  \bibfield  {author} {\bibinfo {author} {\bibfnamefont {{\O}.}~\bibnamefont
  {Gr{\o}n}},\ }\href@noop {} {\bibfield  {journal} {\bibinfo  {journal} {Am.
  J. Phys.}\ }\textbf {\bibinfo {volume} {43}},\ \bibinfo {pages} {869}
  (\bibinfo {year} {1975})}\BibitemShut {NoStop}%
\bibitem [{\citenamefont {Gr{\o}n}(1977)}]{inercial13}%
  \BibitemOpen
  \bibfield  {author} {\bibinfo {author} {\bibfnamefont {{\O}.}~\bibnamefont
  {Gr{\o}n}},\ }\href@noop {} {\bibfield  {journal} {\bibinfo  {journal} {Int.
  J. Theor. Phys.}\ }\textbf {\bibinfo {volume} {16}},\ \bibinfo {pages} {603}
  (\bibinfo {year} {1977})}\BibitemShut {NoStop}%
\bibitem [{\citenamefont {Konno}\ and\ \citenamefont
  {Takahashi}(2012)}]{inercial14}%
  \BibitemOpen
  \bibfield  {author} {\bibinfo {author} {\bibfnamefont {K.}~\bibnamefont
  {Konno}}\ and\ \bibinfo {author} {\bibfnamefont {R.}~\bibnamefont
  {Takahashi}},\ }\href {\doibase 10.1103/PhysRevD.85.061502} {\bibfield
  {journal} {\bibinfo  {journal} {Phys. Rev. D}\ }\textbf {\bibinfo {volume}
  {85}},\ \bibinfo {pages} {061502} (\bibinfo {year} {2012})},\ \Eprint
  {http://arxiv.org/abs/1201.5188} {arXiv:1201.5188 [gr-qc]} \BibitemShut
  {NoStop}%
\bibitem [{\citenamefont {Hosseinpour}\ and\ \citenamefont
  {Hassanabadi}(2015)}]{inercial15}%
  \BibitemOpen
  \bibfield  {author} {\bibinfo {author} {\bibfnamefont {M.}~\bibnamefont
  {Hosseinpour}}\ and\ \bibinfo {author} {\bibfnamefont {H.}~\bibnamefont
  {Hassanabadi}},\ }\href {\doibase 10.1140/epjp/i2015-15236-8} {\bibfield
  {journal} {\bibinfo  {journal} {Eur. Phys. J. Plus}\ }\textbf {\bibinfo
  {volume} {130}},\ \bibinfo {pages} {236} (\bibinfo {year} {2015})},\ \Eprint
  {http://arxiv.org/abs/1505.00096} {arXiv:1505.00096 [hep-th]} \BibitemShut
  {NoStop}%
\bibitem [{\citenamefont {Mota}\ and\ \citenamefont
  {Bakke}(2014)}]{inercial16}%
  \BibitemOpen
  \bibfield  {author} {\bibinfo {author} {\bibfnamefont {H.~F.}\ \bibnamefont
  {Mota}}\ and\ \bibinfo {author} {\bibfnamefont {K.}~\bibnamefont {Bakke}},\
  }\href {\doibase 10.1103/PhysRevD.89.027702} {\bibfield  {journal} {\bibinfo
  {journal} {Phys. Rev. D}\ }\textbf {\bibinfo {volume} {89}},\ \bibinfo
  {pages} {027702} (\bibinfo {year} {2014})},\ \Eprint
  {http://arxiv.org/abs/1401.3728} {arXiv:1401.3728 [hep-th]} \BibitemShut
  {NoStop}%
\bibitem [{\citenamefont {Chowdhury}\ and\ \citenamefont
  {Basu}(2013)}]{inercial17}%
  \BibitemOpen
  \bibfield  {author} {\bibinfo {author} {\bibfnamefont {D.}~\bibnamefont
  {Chowdhury}}\ and\ \bibinfo {author} {\bibfnamefont {B.}~\bibnamefont
  {Basu}},\ }\href {\doibase 10.1016/j.aop.2013.09.011} {\bibfield  {journal}
  {\bibinfo  {journal} {Annals Phys.}\ }\textbf {\bibinfo {volume} {339}},\
  \bibinfo {pages} {358} (\bibinfo {year} {2013})},\ \Eprint
  {http://arxiv.org/abs/1309.1376} {arXiv:1309.1376 [cond-mat.mes-hall]}
  \BibitemShut {NoStop}%
\bibitem [{\citenamefont {Dvornikov}(2015)}]{inercial18}%
  \BibitemOpen
  \bibfield  {author} {\bibinfo {author} {\bibfnamefont {M.}~\bibnamefont
  {Dvornikov}},\ }\href {\doibase 10.1142/S0217732315300177} {\bibfield
  {journal} {\bibinfo  {journal} {Mod. Phys. Lett. A}\ }\textbf {\bibinfo
  {volume} {30}},\ \bibinfo {pages} {1530017} (\bibinfo {year} {2015})},\
  \Eprint {http://arxiv.org/abs/1503.01431} {arXiv:1503.01431 [hep-ph]}
  \BibitemShut {NoStop}%
\bibitem [{\citenamefont {Ambrus}\ and\ \citenamefont
  {Winstanley}(2016)}]{inercial19}%
  \BibitemOpen
  \bibfield  {author} {\bibinfo {author} {\bibfnamefont {V.~E.}\ \bibnamefont
  {Ambrus}}\ and\ \bibinfo {author} {\bibfnamefont {E.}~\bibnamefont
  {Winstanley}},\ }\href {\doibase 10.1103/PhysRevD.93.104014} {\bibfield
  {journal} {\bibinfo  {journal} {Phys. Rev. D}\ }\textbf {\bibinfo {volume}
  {93}},\ \bibinfo {pages} {104014} (\bibinfo {year} {2016})},\ \Eprint
  {http://arxiv.org/abs/1512.05239} {arXiv:1512.05239 [hep-th]} \BibitemShut
  {NoStop}%
\bibitem [{\citenamefont {Bakke}(2010)}]{bakke2}%
  \BibitemOpen
  \bibfield  {author} {\bibinfo {author} {\bibfnamefont {K.}~\bibnamefont
  {Bakke}},\ }\href {\doibase http://dx.doi.org/10.1016/j.physleta.2010.09.046}
  {\bibfield  {journal} {\bibinfo  {journal} {Phys. Lett. A}\ }\textbf
  {\bibinfo {volume} {374}},\ \bibinfo {pages} {4642 } (\bibinfo {year}
  {2010})}\BibitemShut {NoStop}%
\bibitem [{\citenamefont {Bakke}(2013{\natexlab{a}})}]{bakke3}%
  \BibitemOpen
  \bibfield  {author} {\bibinfo {author} {\bibfnamefont {K.}~\bibnamefont
  {Bakke}},\ }\href {\doibase 10.1142/S0217984913500188} {\bibfield  {journal}
  {\bibinfo  {journal} {Modern Physics Letters B}\ }\textbf {\bibinfo {volume}
  {27}},\ \bibinfo {pages} {1350018} (\bibinfo {year}
  {2013}{\natexlab{a}})}\BibitemShut {NoStop}%
\bibitem [{\citenamefont {Mashhoon}(1988)}]{inercial4}%
  \BibitemOpen
  \bibfield  {author} {\bibinfo {author} {\bibfnamefont {B.}~\bibnamefont
  {Mashhoon}},\ }\href@noop {} {\bibfield  {journal} {\bibinfo  {journal}
  {Phys. Rev. Lett.}\ }\textbf {\bibinfo {volume} {61}},\ \bibinfo {pages}
  {2639} (\bibinfo {year} {1988})}\BibitemShut {NoStop}%
\bibitem [{\citenamefont {Santos}\ and\ \citenamefont
  {Barros}(2017{\natexlab{b}})}]{santos2}%
  \BibitemOpen
  \bibfield  {author} {\bibinfo {author} {\bibfnamefont {L.~C.~N.}\
  \bibnamefont {Santos}}\ and\ \bibinfo {author} {\bibfnamefont {C.~C.}\
  \bibnamefont {Barros}},\ }\href {\doibase 10.1140/epjc/s10052-017-4732-x}
  {\bibfield  {journal} {\bibinfo  {journal} {Eur. Phys. J. C}\ }\textbf
  {\bibinfo {volume} {77}},\ \bibinfo {pages} {186} (\bibinfo {year}
  {2017}{\natexlab{b}})}\BibitemShut {NoStop}%
\bibitem [{\citenamefont {Hiscock}(1985)}]{string5}%
  \BibitemOpen
  \bibfield  {author} {\bibinfo {author} {\bibfnamefont {W.~A.}\ \bibnamefont
  {Hiscock}},\ }\href {\doibase 10.1103/PhysRevD.31.3288} {\bibfield  {journal}
  {\bibinfo  {journal} {Phys. Rev. D}\ }\textbf {\bibinfo {volume} {31}},\
  \bibinfo {pages} {3288} (\bibinfo {year} {1985})}\BibitemShut {NoStop}%
\bibitem [{\citenamefont {Bakke}\ and\ \citenamefont {Furtado}(2009)}]{bakke4}%
  \BibitemOpen
  \bibfield  {author} {\bibinfo {author} {\bibfnamefont {K.}~\bibnamefont
  {Bakke}}\ and\ \bibinfo {author} {\bibfnamefont {C.}~\bibnamefont
  {Furtado}},\ }\href {\doibase 10.1103/PhysRevD.80.024033} {\bibfield
  {journal} {\bibinfo  {journal} {Phys. Rev. D}\ }\textbf {\bibinfo {volume}
  {80}},\ \bibinfo {pages} {024033} (\bibinfo {year} {2009})}\BibitemShut
  {NoStop}%
\bibitem [{\citenamefont {Bakke}\ and\ \citenamefont {Furtado}(2010)}]{bakke5}%
  \BibitemOpen
  \bibfield  {author} {\bibinfo {author} {\bibfnamefont {K.}~\bibnamefont
  {Bakke}}\ and\ \bibinfo {author} {\bibfnamefont {C.}~\bibnamefont
  {Furtado}},\ }\href {\doibase 10.1103/PhysRevD.82.084025} {\bibfield
  {journal} {\bibinfo  {journal} {Phys. Rev.}\ }\textbf {\bibinfo {volume}
  {D82}},\ \bibinfo {pages} {084025} (\bibinfo {year} {2010})}\BibitemShut
  {NoStop}%
\bibitem [{\citenamefont {Birrell}\ and\ \citenamefont
  {Davies}(1984)}]{birrel}%
  \BibitemOpen
  \bibfield  {author} {\bibinfo {author} {\bibfnamefont {N.}~\bibnamefont
  {Birrell}}\ and\ \bibinfo {author} {\bibfnamefont {P.}~\bibnamefont
  {Davies}},\ }\href {http://books.google.com.br/books?id=SEnaUnrqzrUC} {\emph
  {\bibinfo {title} {Quantum Fields in Curved Space}}},\ Cambridge Monographs
  on Mathematical Physics\ (\bibinfo  {publisher} {Cambridge University
  Press},\ \bibinfo {year} {1984})\BibitemShut {NoStop}%
\bibitem [{\citenamefont {Cavalcanti~de Oliveira}\ and\ \citenamefont
  {Bezerra~de Mello}(2006)}]{string12}%
  \BibitemOpen
  \bibfield  {author} {\bibinfo {author} {\bibfnamefont {A.~L.}\ \bibnamefont
  {Cavalcanti~de Oliveira}}\ and\ \bibinfo {author} {\bibfnamefont {E.~R.}\
  \bibnamefont {Bezerra~de Mello}},\ }\href {\doibase
  10.1088/0264-9381/23/17/009} {\bibfield  {journal} {\bibinfo  {journal}
  {Class. Quant. Grav.}\ }\textbf {\bibinfo {volume} {23}},\ \bibinfo {pages}
  {5249} (\bibinfo {year} {2006})},\ \Eprint
  {http://arxiv.org/abs/hep-th/0603036} {arXiv:hep-th/0603036 [hep-th]}
  \BibitemShut {NoStop}%
\bibitem [{\citenamefont {Figueiredo~Medeiros}\ and\ \citenamefont
  {de~Mello}(2012)}]{string13}%
  \BibitemOpen
  \bibfield  {author} {\bibinfo {author} {\bibfnamefont {E.~R.}\ \bibnamefont
  {Figueiredo~Medeiros}}\ and\ \bibinfo {author} {\bibfnamefont {E.~R.~B.}\
  \bibnamefont {de~Mello}},\ }\href {\doibase 10.1140/epjc/s10052-012-2051-9}
  {\bibfield  {journal} {\bibinfo  {journal} {Eur. Phys. J.}\ }\textbf
  {\bibinfo {volume} {C72}},\ \bibinfo {pages} {2051} (\bibinfo {year}
  {2012})},\ \Eprint {http://arxiv.org/abs/1108.3786} {arXiv:1108.3786
  [hep-th]} \BibitemShut {NoStop}%
\bibitem [{\citenamefont {Hehl}\ and\ \citenamefont {Ni}(1990)}]{inercial5}%
  \BibitemOpen
  \bibfield  {author} {\bibinfo {author} {\bibfnamefont {F.}~\bibnamefont
  {Hehl}}\ and\ \bibinfo {author} {\bibfnamefont {W.-T.}\ \bibnamefont {Ni}},\
  }\href@noop {} {\bibfield  {journal} {\bibinfo  {journal} {Phys. Rev. D}\
  }\textbf {\bibinfo {volume} {42}},\ \bibinfo {pages} {2045} (\bibinfo {year}
  {1990})}\BibitemShut {NoStop}%
\bibitem [{\citenamefont {Sagnac}(1913{\natexlab{a}})}]{inercial1}%
  \BibitemOpen
  \bibfield  {author} {\bibinfo {author} {\bibfnamefont {G.}~\bibnamefont
  {Sagnac}},\ }\href@noop {} {\bibfield  {journal} {\bibinfo  {journal} {C.R.
  Acad. Sci.}\ }\textbf {\bibinfo {volume} {157}},\ \bibinfo {pages} {708}
  (\bibinfo {year} {1913}{\natexlab{a}})}\BibitemShut {NoStop}%
\bibitem [{\citenamefont {Post}(1967)}]{inercial2}%
  \BibitemOpen
  \bibfield  {author} {\bibinfo {author} {\bibfnamefont {E.}~\bibnamefont
  {Post}},\ }\href@noop {} {\bibfield  {journal} {\bibinfo  {journal} {Rev.
  Mod. Phys.}\ }\textbf {\bibinfo {volume} {39}},\ \bibinfo {pages} {475}
  (\bibinfo {year} {1967})}\BibitemShut {NoStop}%
\bibitem [{\citenamefont {Sagnac}(1913{\natexlab{b}})}]{inercial3}%
  \BibitemOpen
  \bibfield  {author} {\bibinfo {author} {\bibfnamefont {G.}~\bibnamefont
  {Sagnac}},\ }\href@noop {} {\bibfield  {journal} {\bibinfo  {journal} {C. R.
  Acad. Sci. (Paris)}\ }\textbf {\bibinfo {volume} {157}} (\bibinfo {year}
  {1913}{\natexlab{b}})}\BibitemShut {NoStop}%
\bibitem [{\citenamefont {Bruce}\ and\ \citenamefont
  {Minning}(1993)}]{kgoscillator1}%
  \BibitemOpen
  \bibfield  {author} {\bibinfo {author} {\bibfnamefont {S.}~\bibnamefont
  {Bruce}}\ and\ \bibinfo {author} {\bibfnamefont {P.}~\bibnamefont
  {Minning}},\ }\href@noop {} {\bibfield  {journal} {\bibinfo  {journal} {Il
  Nuovo Cimento A}\ }\textbf {\bibinfo {volume} {106}},\ \bibinfo {pages} {711}
  (\bibinfo {year} {1993})}\BibitemShut {NoStop}%
\bibitem [{\citenamefont {Carvalho}\ \emph {et~al.}(2016)\citenamefont
  {Carvalho}, \citenamefont {de~M.~Carvalho}, \citenamefont {Cavalcante},\ and\
  \citenamefont {Furtado}}]{kgoscillator2}%
  \BibitemOpen
  \bibfield  {author} {\bibinfo {author} {\bibfnamefont {J.}~\bibnamefont
  {Carvalho}}, \bibinfo {author} {\bibfnamefont {A.~M.}\ \bibnamefont
  {de~M.~Carvalho}}, \bibinfo {author} {\bibfnamefont {E.}~\bibnamefont
  {Cavalcante}}, \ and\ \bibinfo {author} {\bibfnamefont {C.}~\bibnamefont
  {Furtado}},\ }\href {\doibase 10.1140/epjc/s10052-016-4189-3} {\bibfield
  {journal} {\bibinfo  {journal} {Eur. Phys. J.}\ }\textbf {\bibinfo {volume}
  {C76}},\ \bibinfo {pages} {365} (\bibinfo {year} {2016})},\ \Eprint
  {http://arxiv.org/abs/1603.06292} {arXiv:1603.06292 [hep-th]} \BibitemShut
  {NoStop}%
\bibitem [{\citenamefont {Xiao}, \citenamefont {Long},\ and\ \citenamefont
  {Cai}(2011)}]{kgoscillator3}%
  \BibitemOpen
  \bibfield  {author} {\bibinfo {author} {\bibfnamefont {Y.-J.}\ \bibnamefont
  {Xiao}}, \bibinfo {author} {\bibfnamefont {Z.-W.}\ \bibnamefont {Long}}, \
  and\ \bibinfo {author} {\bibfnamefont {S.-H.}\ \bibnamefont {Cai}},\ }\href
  {\doibase 10.1007/s10773-011-0811-1} {\bibfield  {journal} {\bibinfo
  {journal} {Int. J. Theor. Phys.}\ }\textbf {\bibinfo {volume} {50}},\
  \bibinfo {pages} {3105} (\bibinfo {year} {2011})}\BibitemShut {NoStop}%
\bibitem [{\citenamefont {Jian-Hua}, \citenamefont {Kang},\ and\ \citenamefont
  {Sayipjamal}(2008)}]{kgoscillator4}%
  \BibitemOpen
  \bibfield  {author} {\bibinfo {author} {\bibfnamefont {W.}~\bibnamefont
  {Jian-Hua}}, \bibinfo {author} {\bibfnamefont {L.}~\bibnamefont {Kang}}, \
  and\ \bibinfo {author} {\bibfnamefont {D.}~\bibnamefont {Sayipjamal}},\
  }\href@noop {} {\bibfield  {journal} {\bibinfo  {journal} {Chin. Phys. C}\
  }\textbf {\bibinfo {volume} {32}},\ \bibinfo {pages} {803} (\bibinfo {year}
  {2008})}\BibitemShut {NoStop}%
\bibitem [{\citenamefont {Wen-Chao}(2003)}]{kgoscillator5}%
  \BibitemOpen
  \bibfield  {author} {\bibinfo {author} {\bibfnamefont {Q.}~\bibnamefont
  {Wen-Chao}},\ }\href@noop {} {\bibfield  {journal} {\bibinfo  {journal}
  {Chinese Phys.}\ }\textbf {\bibinfo {volume} {12}} (\bibinfo {year}
  {2003})}\BibitemShut {NoStop}%
\bibitem [{\citenamefont {Bakke}\ and\ \citenamefont
  {Furtado}(2015)}]{kgoscillator6}%
  \BibitemOpen
  \bibfield  {author} {\bibinfo {author} {\bibfnamefont {K.}~\bibnamefont
  {Bakke}}\ and\ \bibinfo {author} {\bibfnamefont {C.}~\bibnamefont
  {Furtado}},\ }\href {\doibase 10.1016/j.aop.2015.01.028} {\bibfield
  {journal} {\bibinfo  {journal} {Annals Phys.}\ }\textbf {\bibinfo {volume}
  {355}},\ \bibinfo {pages} {48} (\bibinfo {year} {2015})},\ \Eprint
  {http://arxiv.org/abs/1411.6988} {arXiv:1411.6988 [quant-ph]} \BibitemShut
  {NoStop}%
\bibitem [{\citenamefont {Maluf}(2011)}]{kgoscillator7}%
  \BibitemOpen
  \bibfield  {author} {\bibinfo {author} {\bibfnamefont {R.~V.}\ \bibnamefont
  {Maluf}},\ }\href {\doibase 10.1142/S0217751X11054887} {\bibfield  {journal}
  {\bibinfo  {journal} {Int. J. Mod. Phys. A}\ }\textbf {\bibinfo {volume}
  {26}},\ \bibinfo {pages} {4991} (\bibinfo {year} {2011})},\ \Eprint
  {http://arxiv.org/abs/1101.2801} {arXiv:1101.2801 [hep-th]} \BibitemShut
  {NoStop}%
\bibitem [{\citenamefont {Vitória}, \citenamefont {Furtado},\ and\
  \citenamefont {Bakke}(2016)}]{kgoscillator8}%
  \BibitemOpen
  \bibfield  {author} {\bibinfo {author} {\bibfnamefont {R.~L.~L.}\
  \bibnamefont {Vitória}}, \bibinfo {author} {\bibfnamefont {C.}~\bibnamefont
  {Furtado}}, \ and\ \bibinfo {author} {\bibfnamefont {K.}~\bibnamefont
  {Bakke}},\ }\href {\doibase 10.1016/j.aop.2016.03.016} {\bibfield  {journal}
  {\bibinfo  {journal} {Annals Phys.}\ }\textbf {\bibinfo {volume} {370}},\
  \bibinfo {pages} {128} (\bibinfo {year} {2016})},\ \Eprint
  {http://arxiv.org/abs/1511.05072} {arXiv:1511.05072 [quant-ph]} \BibitemShut
  {NoStop}%
\bibitem [{\citenamefont {Liang}\ and\ \citenamefont
  {Yang}(2012)}]{kgoscillator9}%
  \BibitemOpen
  \bibfield  {author} {\bibinfo {author} {\bibfnamefont {M.-L.}\ \bibnamefont
  {Liang}}\ and\ \bibinfo {author} {\bibfnamefont {R.-L.}\ \bibnamefont
  {Yang}},\ }\href {\doibase 10.1142/S0217751X12500479} {\bibfield  {journal}
  {\bibinfo  {journal} {Int. J. Mod. Phys. A}\ }\textbf {\bibinfo {volume}
  {27}},\ \bibinfo {pages} {1250047} (\bibinfo {year} {2012})}\BibitemShut
  {NoStop}%
\bibitem [{\citenamefont {Rao}\ and\ \citenamefont
  {Kagali}(2007)}]{kgoscillator10}%
  \BibitemOpen
  \bibfield  {author} {\bibinfo {author} {\bibfnamefont {N.~A.}\ \bibnamefont
  {Rao}}\ and\ \bibinfo {author} {\bibfnamefont {B.}~\bibnamefont {Kagali}},\
  }\href@noop {} {\bibfield  {journal} {\bibinfo  {journal} {Phys. Scripta}\
  }\textbf {\bibinfo {volume} {77}},\ \bibinfo {pages} {015003} (\bibinfo
  {year} {2007})}\BibitemShut {NoStop}%
\bibitem [{\citenamefont {Abramowitz}\ and\ \citenamefont
  {Stegun}(1964)}]{abramo}%
  \BibitemOpen
  \bibfield  {author} {\bibinfo {author} {\bibfnamefont {M.}~\bibnamefont
  {Abramowitz}}\ and\ \bibinfo {author} {\bibfnamefont {I.}~\bibnamefont
  {Stegun}},\ }\href {https://books.google.com.br/books?id=MtU8uP7XMvoC} {\emph
  {\bibinfo {title} {Handbook of Mathematical Functions: With Formulas, Graphs,
  and Mathematical Tables}}},\ Applied mathematics series\ (\bibinfo
  {publisher} {Dover Publications},\ \bibinfo {year} {1964})\BibitemShut
  {NoStop}%
\bibitem [{\citenamefont {Bakke}(2013{\natexlab{b}})}]{inercial10}%
  \BibitemOpen
  \bibfield  {author} {\bibinfo {author} {\bibfnamefont {K.}~\bibnamefont
  {Bakke}},\ }\href {\doibase 10.1007/s10714-013-1561-6} {\bibfield  {journal}
  {\bibinfo  {journal} {Gen.Rel.Grav.}\ }\textbf {\bibinfo {volume} {45}},\
  \bibinfo {pages} {1847} (\bibinfo {year} {2013}{\natexlab{b}})},\ \Eprint
  {http://arxiv.org/abs/1307.2847} {arXiv:1307.2847 [quant-ph]} \BibitemShut
  {NoStop}%
\end{thebibliography}%

\end{document}